# Improvement of accuracy for measurement of 100-km fibre latency with Correlation OTDR

*Florian Azendorf[1], Annika Dochhan[1], Jim Zou[1], Bernhard Schmauss [2], Michael Eiselt[1]*

[1]*ADVA Optical Networking SE, Maerzenquelle 1-3, Meiningen, Germany*
[2]*LHFT, Friedrich-Alexander Universität Erlangen-Nürnberg, 91058 Erlangen, Germany*
*Fazendorf@advaoptical.com*

**Keywords**: Optical time-domain reflectometry, Optical performance monitoring, latency

**Abstract**

We measured the latency of a 100 km fibre link using a Correlation OTDR. Improvements over previous results were achieved by increasing the probe signal rate to 10 Gbit/s, using dispersion compensation gratings, and coupling the receiver time base to an external PPS signal.

## 1 Introduction

The analogue transmission of radio signals over fibre to a phase array antenna requires a very low differential latency between different optical paths. The acceptable differential latency between two optical paths is 3.2 ps for a 26-GHz radio frequency, corresponding to a phase error of 30 degrees. Based on that requirement, an exact measurement of the fibre latency becomes critical for future 5G networks. Changes of the fibre latency are foremost due to changes in the environmental temperature. A typical value for temperature dependent latency, mainly due to a change of the fibre refractive index, is ~7 ppm/K [1]. This means that with a fibre length of 100 km the temperature dependent latency change is approximately 3.5 nanoseconds per degree Celsius, and a continuous monitoring of the fibre latency appears indispensable. Using a Correlation OTDR, we have demonstrated the single-ended (reflective) latency measurement of a 100 km fibre under test (FuT) with an accuracy of better than 11 ps as compared to a two-ended (transmissive) measurement [2]. In this paper, we report on the improvement of the measurement accuracy. In the previous work, a probe signal data rate of 2 Gbit/s was used. To improve the accuracy of determining the reflection point, we increased the data rate to 10 Gbit/s. At this data rate, however, over 200 km propagation distance, chromatic dispersion becomes an issue, broadening the signals reflected from the fibre end. We therefore introduce two chirped fibre Bragg gratings to compensate the fibre chromatic dispersion.

Secondly, the clock used to measure the signal propagation time impacts the accuracy of measuring the absolute fibre latency. The specified clock accuracy of the oscilloscope used in the previous setup was approximately 1 ppm, resulting in a latency time error of 500 ps. We improved the clock accuracy by locking the clock to a pulse per second (PPS) signal from the global navigation satellite system (GNSS) with a specified accuracy of 5 ppb, resulting in a latency error of only 2.5 ps for the 100 km fibre.

## 2 Measurement setup

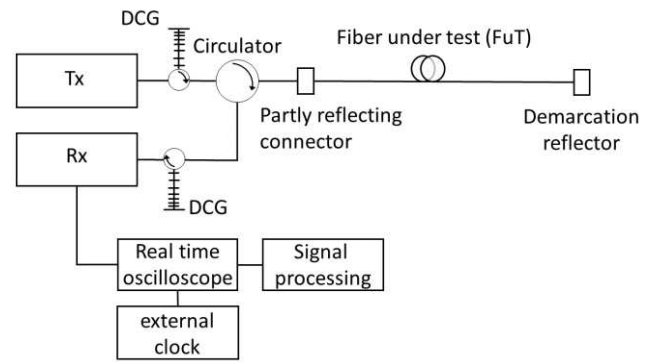

Fig. 1 Schematic of improved measurement setup. DCG: dispersion compensating grating.

A 100-km standard single-mode fibre in a temperature-controlled cabinet was characterized with the setup shown in Fig. 1. A continuous wave laser signal with a wavelength of 1550 nm was modulated, using a Mach-Zehnder modulator, with a test pattern, followed by zeros. The overall packet length was 1 ms, sufficient to measure a 100-km FuT. As test pattern, we used two complementary Golay sequences [3] and, after processing, added the correlation results. To improve the SNR as compared to the previous publication, we doubled the length of the Golay sequences to 4096 bits. The modulated probe signal was sent through a dispersion compensating fibre grating (DCG) with a dispersion of -1700 ps/nm, then fed to the FuT via a circulator. Part of the probe signal was reflected at the fibre input connector, serving as a reference point. A demarcation reflector with a reflection factor of 95% was installed at the fibre output to achieve a higher reflection. The reflected signal was routed by the circulator into a second DCG with a dispersion of -1700 ps/nm and then received with a PIN/TIA combination. The oscilloscope recorded and averaged 4000 signal traces, sampled with 50 GS/s. To improve the timing accuracy, the real time oscilloscope was synchronized with an external oven-controlled crystal oscillator (OCXO) with a stability of 5 ppb. During our experiment, the OCXO was always disciplined by around five GNSS satellites to guarantee the long-term time accuracy. The averaged received signal was correlated with the transmitted respective Golay sequence. In Fig. 2, the averaged received signal, reflected from the fibre end is shown. Due to the fibre round trip loss of 40 dB and the resolution of the oscilloscope of only 7 bits, the signal on/off variation corresponds to much less than a bit and the test pattern cannot be directly observed. After correlation with the transmitted bit sequence, however, the correlation peak of the reflected signal is clearly distinguishable, as shown in Fig. 3. While the sampling rate of 50 GS/s yielded a sample resolution of 20 ps, a raised cosine function was fitted to the correlation values around the highest reflection peak to improve the position accuracy to a few picoseconds, as shown in Fig. 4.

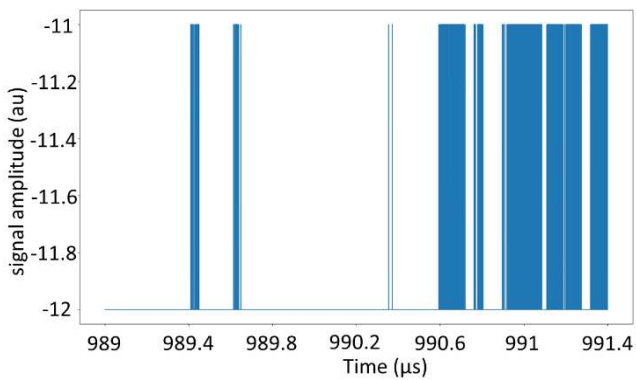

Fig. 2 Received signal from fibre end reflection after averaging on oscilloscope with 7 bits resolution

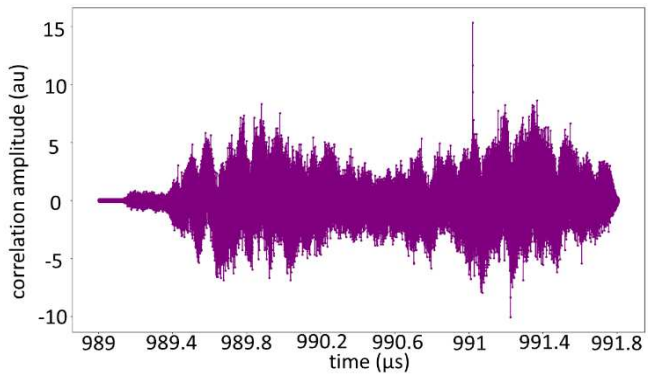

Fig. 3 Averaged signal after correlation with the transmitted Golay sequence, showing the peak from the fibre end reflection with 991 µs round-trip time

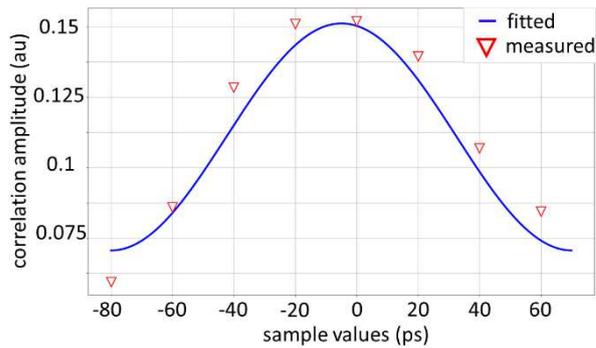

Fig. 4 Fit of a raised cosine function to correlation values around one correlation peak

## 3 Results

*3.1. Accurate time base*

To measure the accuracy improvement from using an external clock, we recorded the external 10-MHz synchronization signal on the oscilloscope using its internal clock as time base. A trace of 100,000 clock periods were recorded, and the duration was measured to be 9.9999756 ms, yielding an internal clock frequency error of +2.44 ppm. The result was verified by recording 200,000 clock cycles with the internal oscilloscope deviating from the external clock by +2.52 ppm. Using the external clock with a specified accuracy of 5 ppb therefore improved the measurement accuracy by a factor of 500.

*3.2. Repeatability*

After synchronization of the real time oscilloscope to the precise clock, the measurement series were started. First, the repeatability of the measurement was tested in series of 24 consecutive measurements, spaced by approximately 4 minutes. During the measurement time of 105 minutes, however, the temperature of the fibre changed slightly (by less than 0.05 K), still leading to a latency variation of more than 100 ps. To estimate the repeatability of the measurement method, we therefore fitted the latency variation to a 4[th] order polynomial and calculated the measured latency values as a deviation from this polynomial trend. Fig. 5 shows the temperature evolution during evening hours with the increasing latency due to temperature increase in the laboratory. The standard deviation of the measured latency differences from the 4th order trendline was calculated as 2.83 ps.

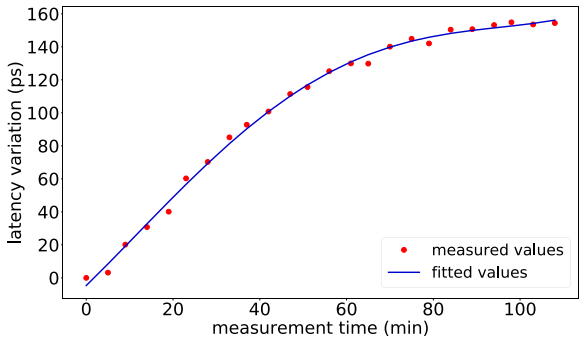

Fig. 5. Fibre latency measurements over 105 minutes in the evening with 4th order polynomial fit.

A more stable temperature in the laboratory was obtained during the morning hours, when the decreasing temperature was reversed by start of the work in the lab. Fig. 6 shows the measured latency values and the 4th order polynomial fit for this series, resulting in a standard deviation of 2.76 ps, comparable to the result of the first series.

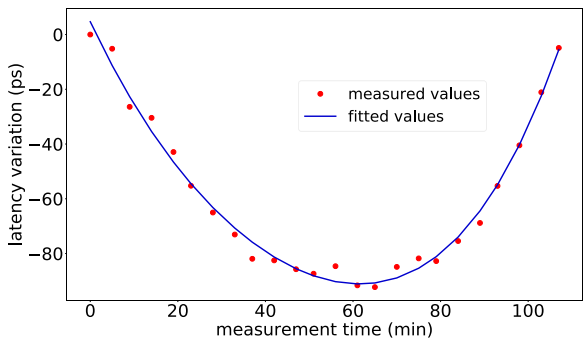

Fig. 6. Fibre latency measurements over 105 minutes in the morning with 4th order polynomial fit.

While the two DCGs introduced before and after the FuT compensated the dispersive broadening of the signal reflected from the fibre end, they caused a broadening of the signal reflected from the fibre input. We measured the correlation width of the input reflection to be 40 ps broader than the end reflection width. Due to the better signal-to-noise ratio of the input reflection, though, this higher pulse width does not impact the accuracy of the reflection peak determination.

*3.3. Reflection and Single Pass*

We then compared the results of the reflection measurements to the measurements of a single pass fibre latency. In a similar setup as in Fig. 1, the signal was sent into to fibre from one end, while the other end was connected to the receiver. As reference for the single pass latency, the FuT was removed and the two DCGs and one reflective connector were connected. Then the FuT was inserted and the combined latency was measured. Alternating, reflective and single pass measurements were performed, separated by approximately four minutes. 4th order polynomials were fitted to the results of each measurement type. Fig. 7 shows the measured latency values and the 4th order polynomial fits for reflective and single pass measurements. The average offset between both curves was 3.9 ps, which can be partly attributed to the different chromatic dispersion experienced by the transmitted and reflected signals. Furthermore, the latency of the demarcation reflector, which was only present in the reflective measurement, was accounted for with a round-trip latency of 160 ps, which could be slightly less than its actual value.

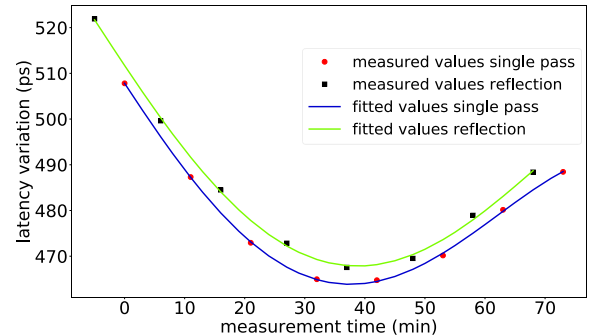

Fig. 7 Comparison of reflective and single pass measurements over 70 minutes in the morning with 4th order polynomial fits to both series

## 4 Conclusion

We improved the correlation OTDR based fibre latency measurement method for long fibres by using a test signal rate of 10 Gbit/s, dispersion compensating gratings, and a precise external clock. The accuracy over a 100 km fibre link, compared to a single-pass measurement, was 3.9 ps. A repeatability of approximately 2.8 ps was achieved. Furthermore, the timing accuracy was improved from 2.5 ppm to 5 ppb. It is therefore possible to use this measurement method to accurately monitor the fibre latency of different optical paths in future networks.

## 5 Acknowledgements

This work has received funding from the European Union´s Horizon 2020 research and innovation programme under grant agreement No 762055 (BlueSpace Project).